\documentclass[english,a4paper,preprint]{revtex4}
\usepackage{graphicx}

\usepackage{color}
\usepackage{dcolumn}
\usepackage{babel}
\usepackage{amsmath}

\begin{document}

\title{Structural and dynamical features of multiple metastable glassy 
states in a colloidal system with competing interactions}

\author{Christian L. Klix$^{1,2}$, C. Patrick Royall$^{1,3}$ 
and Hajime Tanaka$^{1}$\\
$^1$Institute of Industrial Science, University of Tokyo, \\
Meguro-ku,Tokyo 153-8505, Japan. \\
$^2$University of Konstanz, 78457 Konstanz, Germany. \\
$^3$School of Chemistry, University of Bristol, Bristol BS8 1TS, UK.}

\begin{abstract}
Systems in which a short-ranged attraction and long-ranged repulsion compete are intrinsically frustrated, leading their structure and dynamics to be dominated either by mesoscopic order or by metastable disorder. Here we report the latter case in a colloidal system with long-ranged electrostatic repulsions and short-ranged depletion attractions. We find a variety of states exhibiting slow non-diffusive dynamics: a gel, a glassy state of clusters, and a state reminiscent of a Wigner glass. Varying the interactions, we find a continuous crossover between the Wigner and cluster glassy states, and a sharp discontinuous transition between the Wigner glassy state and gel. This difference reflects the fact that dynamic arrest is driven by repulsion for the two glassy states and attraction in the case of the gel. 
\end{abstract}

\maketitle

\section{Introduction}

In general, attractive interactions promote ordered phases or condensates,
whereas long ranged repulsions inhibit this tendency, 
fundamentally redefining the system's free energy leading to
complex phases that break translational and/or rotational symmetry. 
Such mesoscopic ordering occurs
in a very diverse range of materials \cite{seul1995,dagotto2005},
from the pasta phase in neutron stars \cite{horowitz2004}, highly
correlated quantum Hall and strongly correlated electron systems such
as high $T_{c}$ superconductors \cite{emery1993,dagotto2005} to
classical systems \cite{seul1995} such as ferromagnetic films, diblock
copolymers, colloids \cite{sciortino2005,tarzia2006}, and biological
systems.
It has also been suggested 
\cite{sciortino2005,sciortino2004,tarzia2006,tarjus2007}
that competing interactions can also cause frustration,
leading to exotic nonergodic disordered states. In the
above examples, the system is able to relax on mesoscopic lengthscales;
it is rather rare to see metastable disorder at the local level. Such
disordered states were however reported for Laponite suspensions,
where electrostatic repulsions compete with van der Waals attractions,
but the anisotropic particles and interactions make 
the situation rather complex \cite{tanaka2004}.

Spheres with a hard core repulsion 
and an attraction have long provided a model
which captures the essence of atoms and small molecules. 
Short-ranged attractions lead to 
gelation due to arrested phase separation \cite{lu2008}. 
At higher densities both hard-sphere and attractive
glasses are found [Fig. 1(g)] \cite{pham2002},
along with gels \cite{gao2007}. 
Long-range repulsions can lead to glasses at low densities
\cite{sirota1989, zaccarelli2007}, and combined with short-ranged
attractions, the behavior is very rich and complex. Indeed, many
properties of biological materials may be connected, fundamentally,
to a short-range attraction and longer-ranged repulsion due to
electrostatic charging from immersion in an aqueous medium
\cite{semmrich2008}. For example globular proteins
are rather well described as spheres with short-ranged attractions
and long-ranged repulsions \cite{stradner2004}. While most biological
systems are much more complex than spheres with competing interactions,
it is clearly important to understand this seemingly simple addition
to well-studied models of atoms, not least as it offers insight into
transitions between metastable states.

Since competing interactions lead to frustration
between phase separation and homogeneity,
a characteristic lengthscale is often predicted from computer simulation, 
for example, periodic lamellae \cite{tarzia2006}
or low-dimensional clusters of a specific size 
\cite{groenewold2001}. These may then undergo hierarchical
self-organization; in particular, clusters may themselves be implicated
in gelation \cite{kroy2004} and undergo dynamical arrest
to form a `cluster glass' \cite{fernandezToledano2009}. 
Despite recent developments \cite{stradner2004,campbell2005,dibble2006},
experimental work in colloidal systems with competing interactions
has so far found little evidence of periodic structures,
although gels with novel structures \cite{campbell2005,dibble2006}
and low-dimensional clusters \cite{campbell2005, sedgwick2004}
have been seen. 

\section{Experimental}

\begin{figure}
\begin{center}
\includegraphics[width=7.0cm]{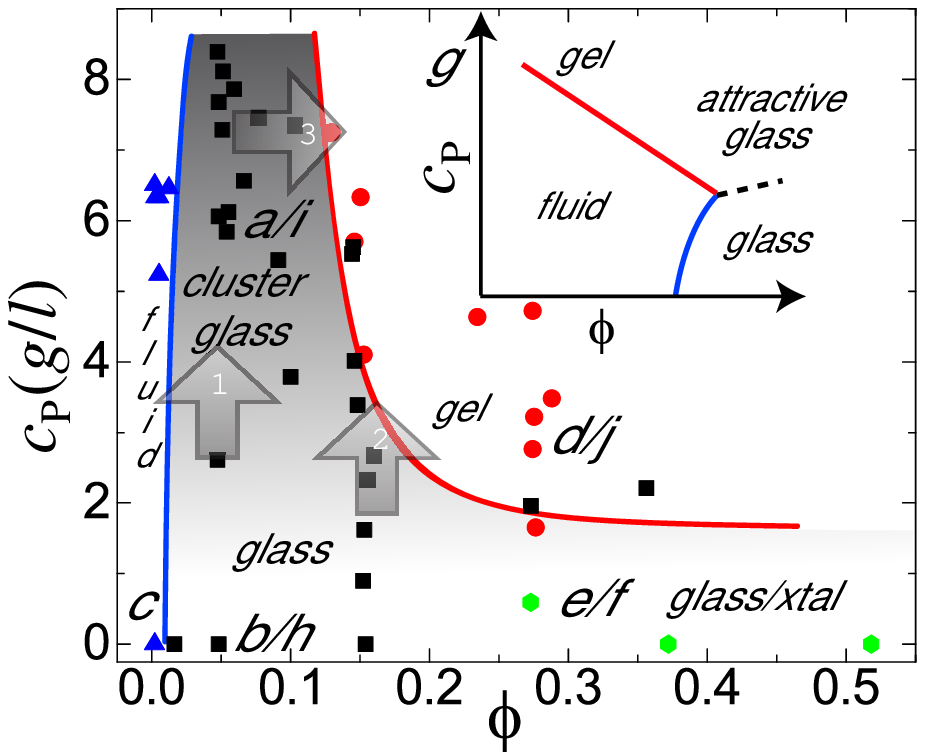}\\
\includegraphics[width=2.5cm]{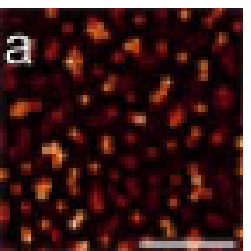}
\hspace{1.0cm}
\includegraphics[width=2.5cm]{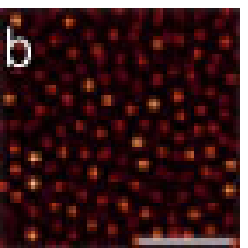}
\hspace{1.0cm}
\includegraphics[width=2.5cm]{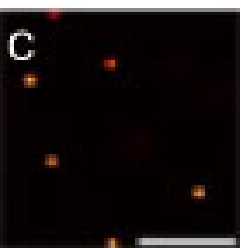}\\
\includegraphics[width=2.5cm]{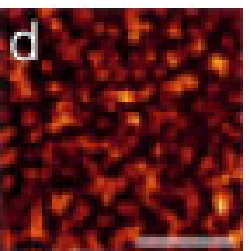}
\hspace{1.0cm} 
\includegraphics[width=2.5cm]{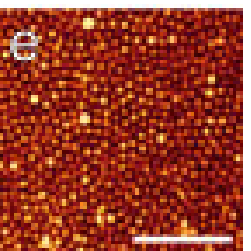}
\hspace{1.0cm}
\includegraphics[width=2.5cm]{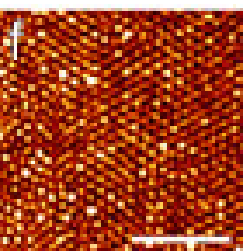}\\
\includegraphics[width=3.0cm]{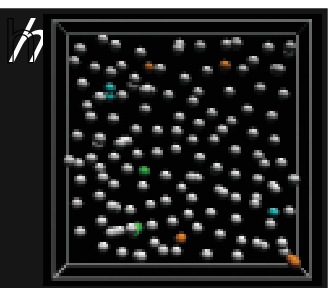}
\hspace{0.5cm} 
\includegraphics[width=3.0cm]{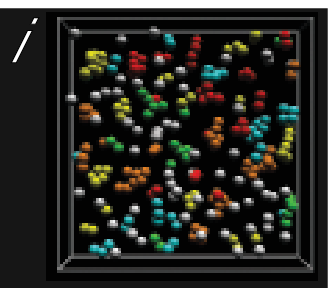}
\hspace{0.5cm}
\includegraphics[width=3.0cm]{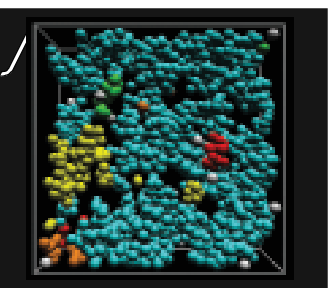}  
\end{center} 
\caption{State diagram of the investigated system and 2D and
3D structures. 
(a) cluster glassy state, $\phi=0.04$, $c_{\rm p}=8.39\:\mathrm{g/l}$. 
(b) Wigner glassy state, $\phi=0.047$, $c_{\rm p}=0$. (c) fluid,
$\phi=0.002$, $c_{\rm p}=0$. (d) gel, $\phi=0.152$,
$c_{\rm p}=4.10$ g/l. (e) Wigner glass, $\phi=0.372$, $c_{\rm p}=0$. 
(f) crystal, $\phi=0.372$, $c_{\rm p}=0$. Bars
are 20 $\mu$m. Boundaries are guides to the eyes. Increasing grey
shading represents increasing amount of clusters. Thick arrows denote
paths 1-3 described in the text. Cluster glassy state is denoted by squares,
gel by circles, crystal/glassy state by hexagons and fluid by triangles.
Grey arrows indicate states for the images (a)-(f).
(g) State diagram of a system without
long-ranged repulsive interactions \cite{pham2002}, showing two glassy 
states at high colloid concentration. 
(h)-(j): 3D structures of (h) Wigner glassy state, 
(i) cluster glassy state, and (j) gel. White particles
have no neighbors, otherwise colors denote connected regions.}
\label{figPhaseDiagram}
\end{figure}

We consider spherical colloids (diameter $\sigma$) 
immersed in a solvent, with a relatively
strong, long-range electrostatic repulsion, and a short-range, tuneable
attraction mediated by non-adsorbing polymer in which the strength
of the attraction is set by the polymer concentration $c_{\rm p}$, and the range by
the polymer size, in this case the polymer-colloid size ratio 
$q \sim 0.19$ (see appendix). 
We determined the
attractions in a very similar system \cite{royall2007}, and the magnitude
of the electrostatic repulsions from fitting the structure of equilibrium
fluids \cite{royall2006} which gave a colloid charge 
number of $Z \sim 600 \pm 200$ (see appendix). 
According to mode-coupling theory for Yukawa systems
\cite{lai1997}, for our parameters we expect a transition to a Wigner
glass at comparable colloid volume fractions to those observed experimentally.
Mixing the samples prior to imaging leads to a randomized
initial state, after which the tuned interactions and colloid volume
fraction $\phi$ lead the system to a (metastable) point on the state diagram
(Fig. \ref{figPhaseDiagram}). 
We sometimes see the formation of a Wigner crystal state at high 
$\phi>0.2$ and low $c_{\rm p}$; 
however, this is a rare event [see Fig. 1(f)] and 
usually the system forms a metastable glassy state before
crystallization can occur (see appendix).  
Using real space structural 
and dynamical analysis, we find
a state diagram dominated by three types of 
glassy states with different disordered structures: Wigner glassy state,
cluster glassy state, and gel. We study
the transitions between these states, and reveal 
the nature of these transitions and its 
link to the interactions leading to slow dynamics.

\section{Results and Discussion}

\begin{figure}
\begin{center}
\includegraphics[width=9.0cm]{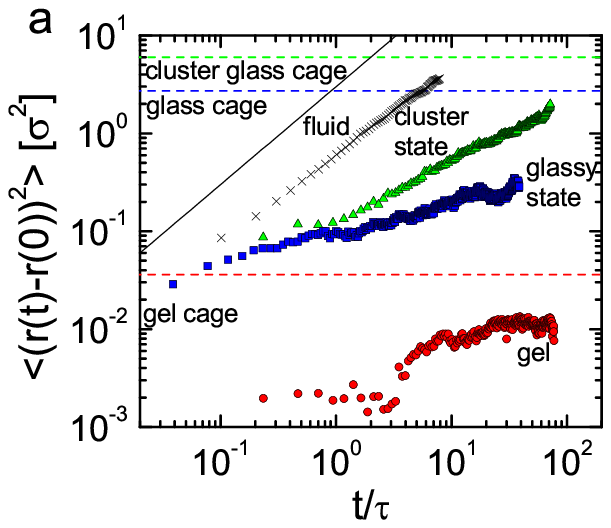} 
\includegraphics[width=9.0cm]{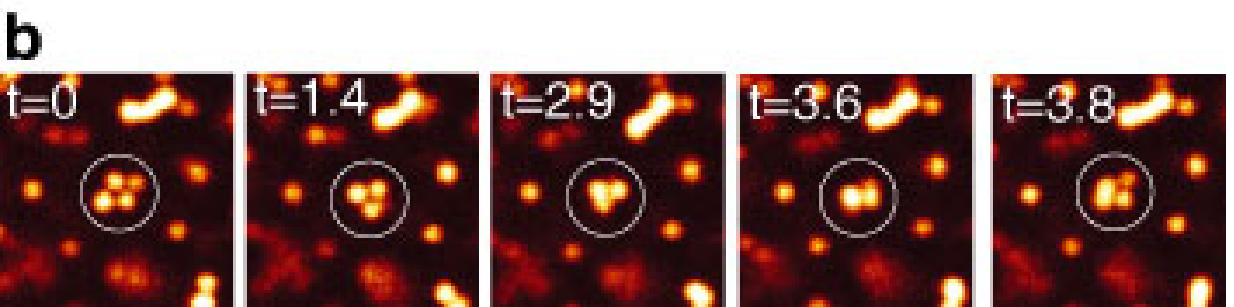} 
\end{center}
\caption{\label{figMSD} (a)  Mean square displacements. 
Crosses=fluid ($\phi=0.002$, $c_{\rm p}=0$), green triangles=cluster glassy 
state ($\phi=0.051$, $c_{\rm p}=5.16$ g/l), red circles=gel 
($\phi=0.275$, $c_{\rm p}=4.88$ g/l) and 
blue squares=glassy state ($\phi=0.069$, $c_{\rm p}=0$). 
$\tau$=characteristic
diffusion time (see appendix). Dashed lines corresponding
to the cage size are plotted for gel ($0.19\sigma$), Wigner state
($1.9\sigma$) and cluster state ($2.1\sigma$), respectively. (b)
Confocal microscope images of a cluster spinning
in it's `cage'). 
Time unit=$\tau$, image width=$19.5\:\mathrm{\mu m}$.}
\end{figure}

We begin by presenting the state diagram in Fig. \ref{figPhaseDiagram},
which underlines the extent to which the system is dominated by dynamically
arrested states. A low-density colloidal fluid ($\phi=0.002$) where the system is ergodic is shown 
in Fig. \ref{figPhaseDiagram}(c). 
Increasing the volume fraction to $\phi=0.016$ results in a glassy  
state where the slow dynamics is driven by the long-ranged electrostatic 
repulsions [Figs. \ref{figPhaseDiagram}(b), (e), and (h)]. 
We thus term this state a
`Wigner glassy state'. At low $\phi$ and higher $c_{\rm p}$, 
we see the formation of clusters and term this state 
a `cluster glassy state' [Figs. \ref{figPhaseDiagram}(a) and (i)].  
Meanwhile,
increasing both $c_{\rm p}$ and $\phi$ results in a gel 
which we define through percolation [Figs. \ref{figPhaseDiagram}(d) and (j)], 
and appears dynamically arrested [Fig. \ref{figMSD}(a)]. 
A typical state diagram for a colloidal system without electrostatic 
repulsions is schematically shown in Fig. 1(g). We see  
striking differences in the structure of the state diagram 
between systems with and without electrostatic repulsions (see also appendix). 
With electrostatics the ergodic region is very 
narrow, around $\phi \sim 0$, and the glassy 
states dominate the state diagram. 
There are three glassy states with very different structures: 
Wigner and cluster glassy states and a gel state. 
We also found three types of transitions between these 
states accompanied by local structural changes. 

Before discussing the structure of these three states in more detail,
let us consider the dynamics. Mean squared displacement (MSD) measurements,
in which the colloids are tracked in two dimensions (2D), are shown
in Fig. \ref{figMSD}. What is clear is that the (ergodic) fluid we see at 
low $\phi$ appears to
exhibit diffusive behavior, and the gel appears dynamically arrested,
within the accuracy to which we can track the particles ($100$
nm or $\sigma/20$). 
The other states show extensive non-diffusive behavior,  
yet they do not reach a clear plateau on the experimental timescale 
although we track the particles for up to 20 hours for the cluster glassy  
state and up to 2 hours for the Wigner glassy state. 
We are limited in particle tracking at long times due to particle loss
(particles diffusing out of our volume), bleaching and drift, along with
residual sedimentation.
We estimate the cage size as the width of the first and second peaks
in $g(r)$ for the Wigner and cluster glassy states
respectively [see Fig. 3(a)]. 
In the case of the gel the cage size is taken as the bond length,
$0.19\sigma$. Concerning the cluster glassy 
state, similar sub-diffusive 
behavior has recently been seen in computer simulation 
\cite{fernandezToledano2009}. 
We note that such behavior is expected from 
the fact that clusters can rotate as rigid bodies [see Fig. 2(b)]. 
This strong decoupling between translational and rotational motion 
may be unique to a cluster glassy state. 
In the case of the Wigner glassy state, based upon simulation work
in a similar system \cite{zaccarelli2008}, we believe that 
our experimental timescales
limit full access to relaxation phenomena. In any
case, the cage size is not reached [see Fig. 2(a)]. 
The soft nature of the interactions is another factor 
which prolongs the sub-diffusive regime and thus makes 
a rigorous confirmation of nonergodicity very challenging. 
The non-diffusive behavior is suggestive of the glassiness of 
these states. Since we do not find a clear plateau experimentally, 
we term these `cluster glassy state' and `Wigner glassy state', respectively, 
although the states may be regarded as nonergodic 
at least practically. 

\begin{figure}
\begin{center}
\includegraphics[width=3.5 cm]{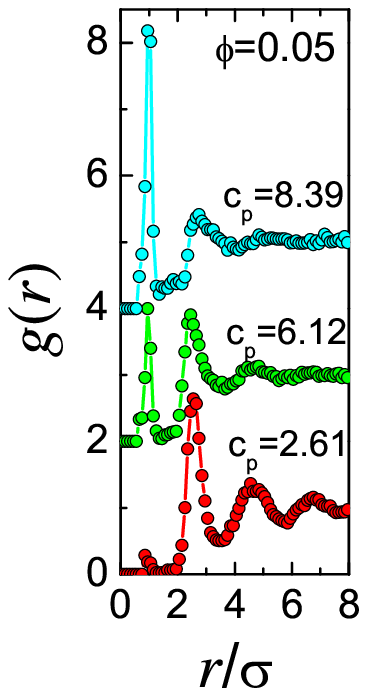}
\includegraphics[width=2.55 cm]{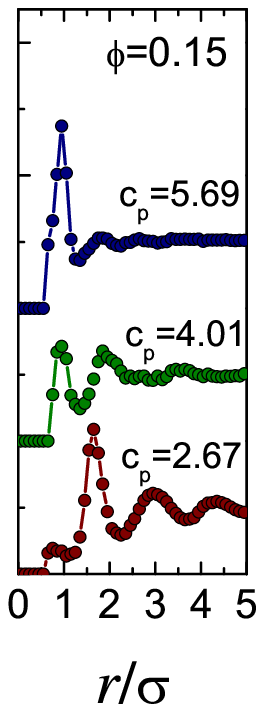}
\includegraphics[width=3.5 cm]{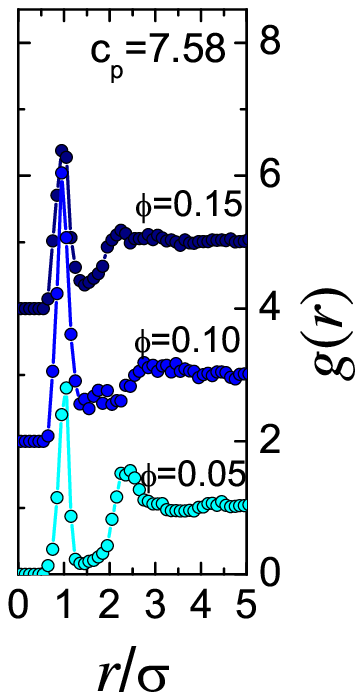}
\includegraphics[width=10.0 cm]{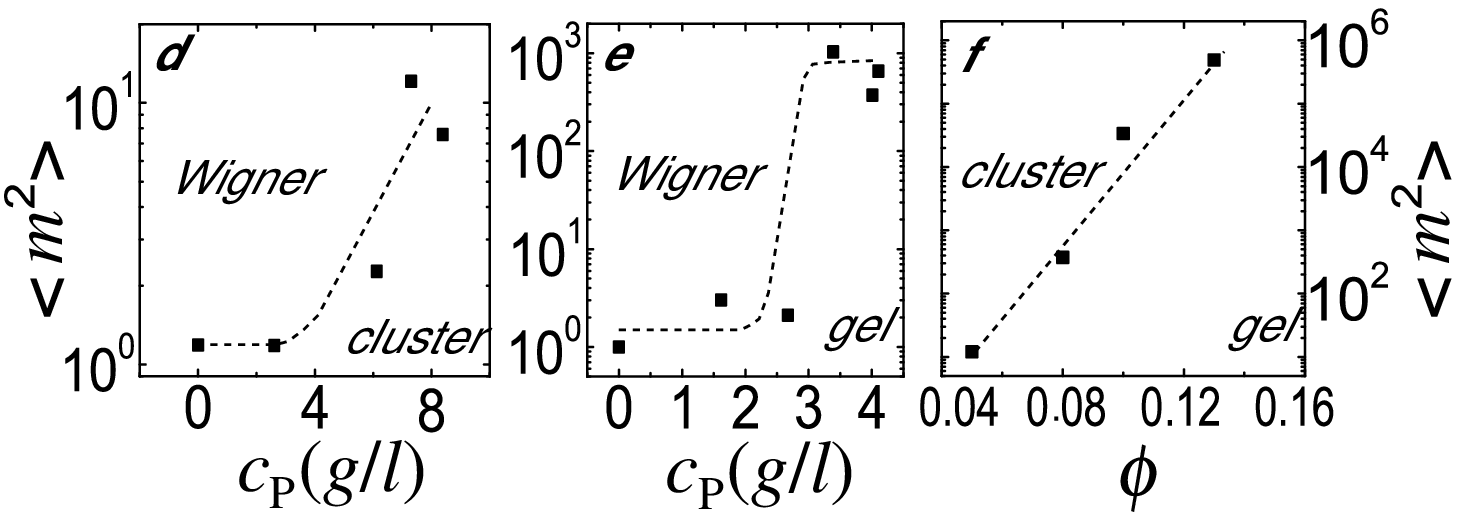} 
\par\end{center}
\caption{
Dependences of 
$g(r)$ and 
$\langle m^{2}\rangle$ on $\phi$ and $c_{\rm p}$. (a) (path 1) shows
the change in $g(r)$ across 
the Wigner-cluster glassy state transition.  
Vertical solid line indicates 
the peak at contact ($r=$$\sigma$), and dashed line indicates 
the first peak of $g(r)$ of the Wigner glassy state, corresponding 
to the average interparticle distance, and the second
peak of $g(r)$ of the cluster glassy state, 
corresponding to the first shell of clusters.  
(b) (path 2) from a Wigner glassy to a gel state, and (c) (path 3) from a 
cluster glassy to a gel state. $c_{\rm p}$ is given in units of g/l. 
Data offset for clarity. (d) 
The change in $\langle m^{2}\rangle$ along path 1 (at $\phi=0.051$), 
(e) along path 2 (at $\phi=0.151$), and along (f) path 3 
(at $c_{\rm p}=7.58$ g/l).
Dashed lines and shading are guides to the eye. 
\label{figG}}
\end{figure}

The state diagram dominated by these three glassy states illustrated
in Fig. \ref{figPhaseDiagram} yields three transitions. 
It is a commonly held view that dynamical arrest is accompanied
by little structural change. While literature on transitions between
glassy states is scarce, the structural change in the attractive-repulsive
glass transition at high packing fraction 
is rather subtle \cite{simeonova2006}.
Conversely, all states identified here are characterized by their
structures [see Figs. 1(h), (i), and (j)]. 
Our analysis shows a considerable variety of metastable structures 
in the response to small changes in parameters. 

Let us now enquire as to the nature of the transitions between these
glassy states. The transition between Wigner and cluster glassy states 
(path 1 in
Fig. \ref{figPhaseDiagram}) is considered in more detail in Figs.
\ref{figG}(a) and (d), which show the radial distribution
function $g(r)$ and the variance in the cluster size distribution 
$\langle m^2 \rangle$, where $m$ is the number of particles per 
cluster. Rather than a sharp transition, the Wigner glassy state 
is unaffected by weak attractions; until the polymer concentration
exceeds around $c_{\rm p}=4$ g/l there is no response in the size distribution.
At higher $c_{\rm p}$, there is an increase in (cluster) size, yet
our data suggest that it occurs rather gradually, {\emph{i.e.}},
passing from the Wigner to the cluster glassy state is a crossover 
rather than a
sharp transition: particles start to form small clusters above a 
certain threshold $c_{\rm p}$ and their size gradually increases with an 
increase in $c_{\rm p}$. The gradual development of the peak at contact 
(at $r=\sigma$) in
the $g(r)$ [Fig. \ref{figG}(a)] further supports this
observation. 
Note that the emergence of the peak at contact is a direct consequence 
of attractions. 
In equilibrium, the repulsive Wigner state (comprised of monomers)
is expected to be a crystal. The same can apply to states of 
\emph{monodisperse}
clusters \cite{mladek2006}, however here the clusters are very polydisperse
[see Fig. \ref{figPhaseDiagram}(a) and appendix], which leads to self-generated 
disorder, and intrinsically suppresses crystallization. 
This source of disorder is
a consequence of a two-level organizational hierarchy, of colloids
forming clusters and then clusters forming a glass which may be characteristic
of a system of competing interactions with different lengthscales.

Proceeding to the transition between the Wigner glassy and gel states 
(path 2 in Fig. \ref{figPhaseDiagram}),
we find a rather different scenario. Raising the colloid volume fraction
to $\phi=0.15$, the $g(r)$ [Fig. 3(b)] 
again shows the development of a peak at contact ($r=\sigma$) 
around $c_{\rm p}=4$ g/l. We recall that at lower
$\phi$, at a similar polymer concentration, clusters began 
to form [Fig. \ref{figG}(a)].
For $\phi=0.15$ this yields percolation 
[Fig. \ref{figPhaseDiagram}(j)], and a sharp transition
to a gel state [Fig. 3(e)], accompanying a strong increase in $\langle m^2 
\rangle$ by about three orders of magnitude. 
This is markedly different from the gradual continuous increase 
in $\langle m^2 \rangle$ for the Wigner-cluster glassy states 
[see Fig. 3(d)].

What happens in the case of the transition between cluster glassy 
and gel states? Path
3 in Fig. \ref{figPhaseDiagram} is shown in Figs. \ref{figG}(c) 
and (f). Unlike the previous transitions, here the polymer
concentration is fixed at $c_{\rm p}=7.4$ g/l. In fact rather little
change in the local structure is seen in the $g(r)$ analysis {[}Fig.
\ref{figG}(c)], and the variance in the cluster size increases
continuously. Our results indicate that, in contrast to paths 1 and 2 in
Fig. \ref{figPhaseDiagram} the cluster glass-gel boundary is delineated
by a percolation transition, rather than by local structural changes.

\begin{figure}
\begin{center}
\includegraphics[width=6.0cm]{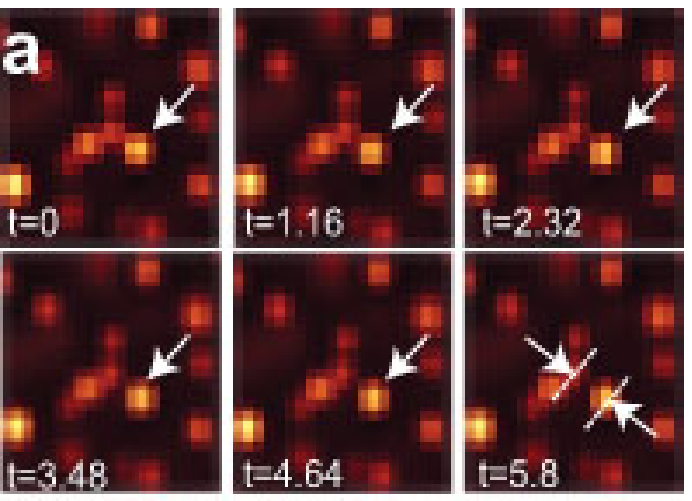}
\includegraphics[width=4.0cm]{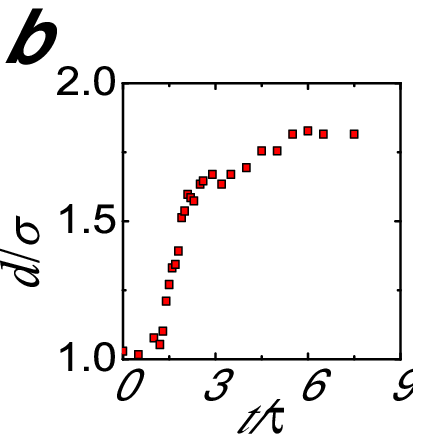} 
\end{center}
\caption{\label{figSpittingOut} Aging mechanism of a 
cluster glassy state. (a) An emission process  
from an $m=5$ to $m=4$ cluster,
as shown by arrows. 
Time $t$ is expressed in units of $\tau/1000$ 
and the image width is $14\:\mathrm{\mu m}$. 
(b) Separation of the emitted particle and cluster, 
as defined by arrows in (a), as a function of $t$. }
\end{figure}

We also observed a novel aging mechanism in the cluster state. While
aging is usually thought of in dynamic terms, here we present a structural
mechanism for aging. This concerns the transition from a state of
high potential energy, a tightly bound cluster of charged particles,
to a smaller cluster, via cluster fission. We never observed any cluster
fusion, and particle tracking shows a continuous rise in the population
of small clusters as a function of time. The emission process occurs
in less than $1/100$ of the characteristic diffusion time, a much
faster timescale than structural relaxation even in the absence
of slow dynamics. This phenomenon is illustrated in Figs. 
\ref{figSpittingOut}(a) and (b).
This cluster state suffers from a complex minimization problem involving
the spatial distributions of colloids, counterions and polymers. The
coupling between these three variables makes the potential energy
landscape rather complex. This fission process allows us to directly
observe a kinetic path from one local minimum to a neighboring minimum
with a lower energy. Details of the cluster distribution and 
its time-evolution are available in the appendix.

Some comments on the colloidal interactions are in order. We have
determined the colloid charge at low $\phi$ and $c_{\rm p}=0$. 
Using the parameters we calculate for the colloid charge, 
for this system we expect a potential at contact of several hundred 
$k_{\rm B}T$,
substantially in excess of the attractive forces mediated by the depletion 
attraction, $\sim 30$ $k_{\rm B}T$ for $c_{\rm p}=4$ $gl^{-1}$ at which we 
find clustering and gelation (see appendix).
We believe that the charge is associated with the free surfaces of 
the colloids, and that clusters may have a markedly lower charge 
per colloid than do the isolated particles \cite{klix2009}.
Calculations show that for such anisotropic charge distributions, 
the electrostatic repulsion
is much reduced, enabling the clustering and gelation 
we observe \cite{hoffman2004}. 

\section{Conclusions}

In closing, we found that the introduction of relatively 
strong, long-ranged repulsions to a colloidal dispersion 
with short-range attractions,
generates
novel glassy states, such as the cluster glassy state, 
and drastically transforms
behavior, at that most fundamental of levels: the ability of the system
to relax locally.  
For a system of short-ranged attractive interactions, without strong,
long-ranged repulsions, it is possible to distinguish the
repulsive hard-sphere glass and the attractive glass, as depicted
in Fig. \ref{figPhaseDiagram}(g) \cite{pham2002},
but there the change in the structure 
is rather subtle \cite{simeonova2006}. 
In contrast, the transitions between cluster glassy and gel states and 
between Wigner glassy and gel states both accompany significant structural
changes from single particle to compact clusters and a percolating
network, respectively. 
The observed behavior can be interpretted as the interplay of 
a Wigner glassy state dominated by the long-range repulsions 
with formation of clusters driven by short-range attractions 
\cite{groenewold2001} and their percolation for gelation. 
Like the neutral system, there appear to be
two drivers of arrest: the gel is driven by attractions and
both Wigner and cluster glassy states are driven by repulsion. 
The effect of long-ranged repulsions on the 
state diagram of a system with short-ranged attractions is an interesting
fundamental problem; at intermediate 
repulsions a cluster fluid emerges \cite{sedgwick2004,campbell2005,dibble2006}.
Since protein solutions are also understood to exhibit comparable
interactions, we conclude that it is at least reasonable to suppose
that equilibrium cluster phases may be found \cite{stradner2004,shukla2008},
and perhaps even cluster glasses. 

The authors are grateful to Didi Derks, Rob Jack and Matthias Schmidt
for a critical reading of the manuscript. 
CLK thanks the Deutscher Akademischer Austauschdienst for financial
assistance. CPR acknowledges the Royal Society for financial support.
HT acknowledges a Grant-in-Aid from the Ministry
of Education, Culture, Sports, Science and Technology, Japan.


\section*{Appendix}


\noindent \textbf{Samples and experimental methods.}
We used poly(methyl methacrylate) (PMMA) colloids sterically stabilized
with polyhydroxyl steric acid. The colloids were labelled with the
fluorescent dye rhodamine isothiocyanate and had a diameter $\sigma=1.95$
$\mu$m with around 5\% polydispersity. Solvents and polymers were
used as received. The polymer used was polystyrene, with a molecular
weight of $M_{\rm w}=2.06\times10^{7}$.  
To closely match the colloid density and
refractive index we used a solvent mixture of cis-decalin and cyclohexyl
bromide. Due to the refractive index matching, the van der Waals interactions
are reduced to a fraction of the thermal energy $k_{\rm B}T$ 
($k_{\rm B}$: Boltzmann's constant; $T$: temperature)  and neglected.
We confined the samples to thin (100 $\mu$m) capillaries which provided
access to the entire sample, but confirmed, with experiments using
larger 500 $\mu$m cells, that this confinement has little effect upon
the behavior. Prior to each experiment, sample vials were washed with
copious quantities of ethanol, and carefully dried. Dry colloids were
then added, and dispersed in solvent, with the polymer solution added
last. The data was collected on a Leica SP5 confocal microscope, fitted
with a resonant scanner. We determined the co-ordinates of each particle
with a precision of around 100 nm \cite{royall2007}. Connectivity
was determined by setting the bond length equal to the sum of the
range of the attractive interactions and the tracking error. Our conclusions
are insensitive to the exact value of the bond length. The time for
a colloid to diffuse one radius is given by the Stokes-Einstein relation
as $\tau=3\pi\eta\sigma^{3}/(4k_{\rm B}T)$ where $\eta$ is the viscosity.
This we measure independently as a function of polymer concentration
using a Rheologica Instruments Stress Tech rheometer.

\vspace{0.5cm}
\noindent \textbf{Determining the interactions between the colloids.}
The behavior we have observed is driven by competing attractive and
repulsive interactions. We shall begin our discussion of the potentials
by considering the attractive depletion interaction. We have used
a polymer of $M_{\rm w}=2.06\times10^{7}$. 
Here we used a `good' solvent, 
and in detailed studies using the same solvent-polymer
mixture, we found good agreement with the Asakura-Oosawa (AO) theory 
\cite{asakura1954},
assuming polymer swelling such that the radius of gyration $R_{\rm G}$
is increased by 35\%. We therefore assumed a similar degree of swelling
here, such that $R_{\rm G}\approx190$ nm, leading to a polymer-colloid
size ratio $q=2R_{\rm G}/\sigma\approx0.19$, which sets the range of
the attractive interaction. 
For the dilute polymer regime, little deviation is expected from AO theory
\cite{asakura1954}, even in the case of some polymer swelling (non-ideality)
\cite{louis2002}. Furthermore, although we add sufficient polymer
that some of our samples are just into the semi-dilute regime, relatively
little deviation is expected from AO theory for these parameters 
\cite{louis2002}.
At the level of this work, therefore, we treat the attractions with
AO theory which leads to a pair interaction between two hard colloidal
spheres in a solution of ideal polymers which reads
\begin{equation}
\beta u_{AO}(r)=\begin{cases}
\text{$\infty$} & \text{for $r<\sigma$}\\
\text{$\frac{\pi(2R_{\rm G})^{3}z_{\rm PR}}{6}\frac{(1+q)^{3}}{q^{3}}$}\\
\text{$\times\{1-\frac{3r}{2(1+q)\sigma}+\frac{r^{3}}{2(1+q)^{3}\sigma^{3}}\}$} & \text{for $r\ge\sigma<\sigma+(2R_{\rm G})$}\\
0 & \text{for $r\ge\sigma+(2R_{\rm G})$}\end{cases}\label{eqAO}
\end{equation}
where $\beta$ is $1/k_{\rm B}T$. The polymer fugacity $z_{\rm PR}$ is
equal to the number density $\rho_{\rm PR}$ of ideal polymers in a reservoir
at the same chemical potential as the colloid-polymer mixture. 
This corresponds to a contact potential of 7.0 $k_{\rm B}T$
per $\mathrm{g/l}$ of polymer reservoir concentration.

Determining the electrostatic interactions in non-aqueous colloidal
dispersions is a non-trivial task \cite{royall2003,royall2007}, as
the charge and ionic strength are very small, although they still
lead to significant interactions. The interactions have nevertheless
been found to be well described by the screened Coulomb or Yukawa interaction,
which reads
\begin{equation}
\beta u_{YUK}(r)=\begin{cases}
\text{$\infty$} & \text{for $r<\sigma$}\\
\text{$\beta\epsilon\frac{\exp(-\kappa(r-\sigma))}{r/\sigma}$} & \text{for $r\ge\sigma$}\end{cases}\label{eqYuk}\end{equation}
where $r$ is the center to center separation of the two
colloids. The contact potential is given by 
\begin{equation}
\beta\epsilon=\frac{Z^{2}}{(1+\kappa\sigma/2)^{2}}\frac{l_{\rm B}}{\sigma}, 
\end{equation}
 where $Z$ is the colloid charge, $\kappa$ is the inverse Debye
screening length and $l_{\rm B}$ is the Bjerrum length.

One may thus assume interactions between the colloids take the 
form of a short-range
`depletion' attraction and long-ranged screened Coulomb repulsion,
and a nearly hard core. Both have been measured in nearly identical
systems and found to be in good agreement with theory 
\cite{royall2007,royall2006}.
Here we have determined the colloid charge and ionic strength by comparison
of the measured $g(r)$ with one obtained from Monte Carlo simulation
according to a Yukawa potential
\cite{royall2006} in the ergodic fluid part of the state diagram
($\phi=4.8\times10^{-4}\pm2\times10^{-5}$, $c_{\rm p}=0$) to be
$Z\approx600\pm200$ $e$ where $e$ is the elementary unit of charge,
and the inverse Debye length $\kappa\sigma\approx1\pm0.4$. The 
radial distribution function $g(r)$
fitting is shown in SFig. 1. This method is found to be consistent
with electrophoretic measurements \cite{royall2006}. 
The values we obtain for the colloid charge are somewhat
(up to 25\%) higher than those quoted previously in literature where 
samples were prepared using 
dry colloids \cite{royall2006, royall2003}. For these parameters 
mode-coupling theory predicts vitrification at higher volume fractions
 \cite{lai1997}. However, this analysis poses a significant
question: the contact potential between two such colloids is 
$\beta\epsilon \sim1000$. This is much greater than  
the strength of the attraction induced by the polymer. 
Throughout the measurements 
we find that clustering of gelation occurs at a polymer concentration 
around $c_{\rm p}=4.0$ gl$^{-1}$, which corresponds to a contact potential of
around 30 $k_{\rm B}T$ [Eq. (\ref{eqAO})]. 

We are reasonably confident about our understanding
of the attraction induced by the polymer \cite{royall2007}. We speculate 
that the charge is in fact different between the isolated colloids upon which measurements of
the charge $Z$ are based
and colloids found in clusters, consistent with behavior we observed
in a similar cluster-forming system \cite{klix2009} and
values quoted in the literature \cite{stradner2004,campbell2005,dibble2006}.
Alternatively, non-pairwise additivity with these long-ranged repulsions
leading to a reduction in apparent repulsion may play an important role
\cite{merrill2009}.
In all of these cases, the quoted values for 
the repulsive interactions 
are stronger than the depletion induced attractions.

Pointers to an understanding of the underlying cause for this discrepancy 
may be found in measurements that indicate a reduction in charge 
per colloid upon clustering \cite{klix2009}. 
Moreover, if this charge were inhomogeneously distributed, a further 
reduction in repulsions is to be 
expected \cite{hoffman2004} upon which clustering due to 
polymer-induced attractions is not unreasonable.
Further work is in progress to elucidate the charging mechanism, 
and possible charge anisotropy, however, 
we note that the weak levels of electrostatic charging in 
these apolar systems severely hamper 
a full understanding.


However, 
the strength of the depletion interaction, as gauged by its strength at 
contact [Eq. (\ref{eqAO})] is rather stronger here than in systems where 
the charge is screened 
\cite{royall2008gel, lu2008}. Comparing the minimum
polymer concentration required to form a gel, we find around a contact 
potential of 
$\beta u_{AO} \approx 11$. When the electrostatic repulsions are screened
one finds $\beta u_{AO} \approx 3$ for 
attractions of comparable range (q=0.18) \cite{royall2008gel}, 
rising slightly to $\beta u_{AO} \approx 5$ for shorter-ranged attractions
\cite{lu2008}. Previous work on charged systems with a much short-ranged 
attraction ($q=0.04$) suggests a threefold increase in the 
contact potential $\beta u_{AO}$
required for gelation \cite{dibble2006} relative to systems where the 
charge is screened.

\vspace{0.5cm}
\centerline{\includegraphics[width=0.5\textwidth]{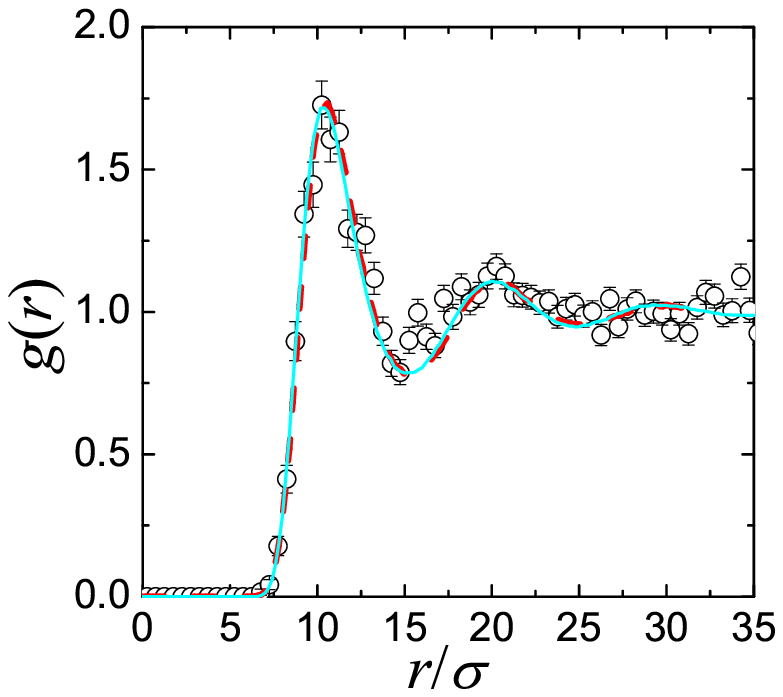}}
\noindent \textbf{SFig. 1: Analysis of g(r). }
Radial distribution
function. Dashed red line corresponds to a colloid charge $Z=400$,
solid cyan to $Z=800$. We assumed that the Debye screening length was
dominated by the colloidal counterions, in other words the system is close to the salt-free limit. 
This leads to a fitting which depends solely upon $Z$.
Lower values of $Z$ gave poor fits, higher values of $Z$ led to crystallization.

\vspace{0.5cm}
\centerline{\includegraphics[width=0.4\textwidth]{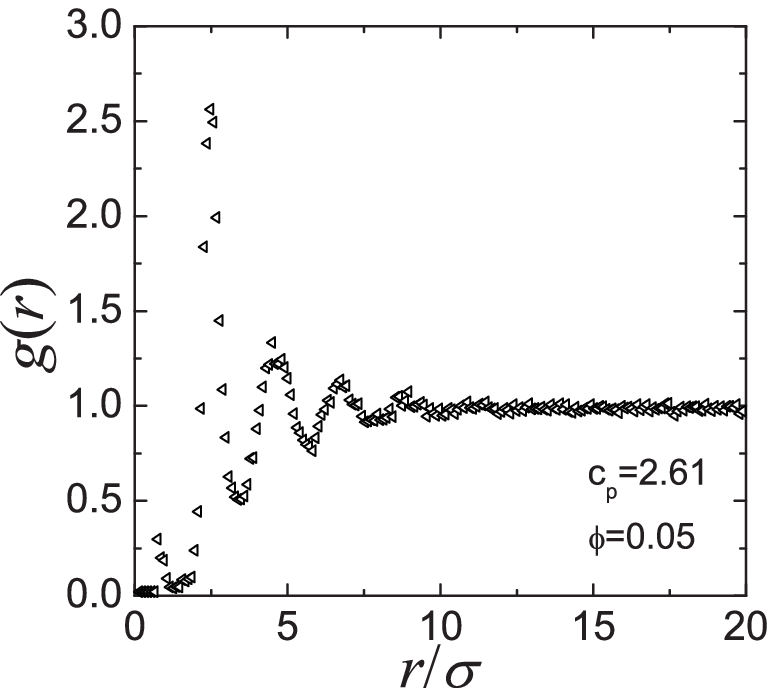}}
\noindent \textbf{SFig. 2: g(r) for repulsive glass.} Decay
of the spatial density correlation at long range suggests an absence
of crystallization and that the state is a repulsive Wigner glass.
Polymer concentration $c_{\rm p}$ is given in units of g/l. 


\vspace{1cm}
\noindent \textbf{On crystallization.}
Occasionally we found that, rather than
a Wigner glassy state [Fig. 1(e)], the system in fact
crystallized [Fig. 1(f)]. That similar samples
may either crystallize or vitrify suggests multiple paths through
the free-energy landscape, as found in some clay suspensions 
\cite{jabbari2007}. Alternatively, the systems size we study
allow the possibility of a variation in nucleation time
between different experiments, leading to the observation
of crystallization in a limited set of samples.

The fact that the crystal forms at all also suggests that the increase
in $\phi$ may promote ordering and reduce the glass-forming ability
of the system. We never observed an ordered phase for $\phi<0.2$,
this is further supported by the radial distribution function in SFig. 2 
which shows an absence of long-ranged order for the repulsive
glass. That we only found a Wigner glass at low $\phi$, an apparently
shallower quench, but occasionally a crystal at higher $\phi$, is
interesting, and may be explained by a stronger screening of the electrostatic
repulsion due to a higher concentration of counter ions at higher
$\phi$, leading to `harder' interactions which promote crystallization. 

\vspace{0.5cm}
\centerline{\includegraphics[width=0.8\textwidth]{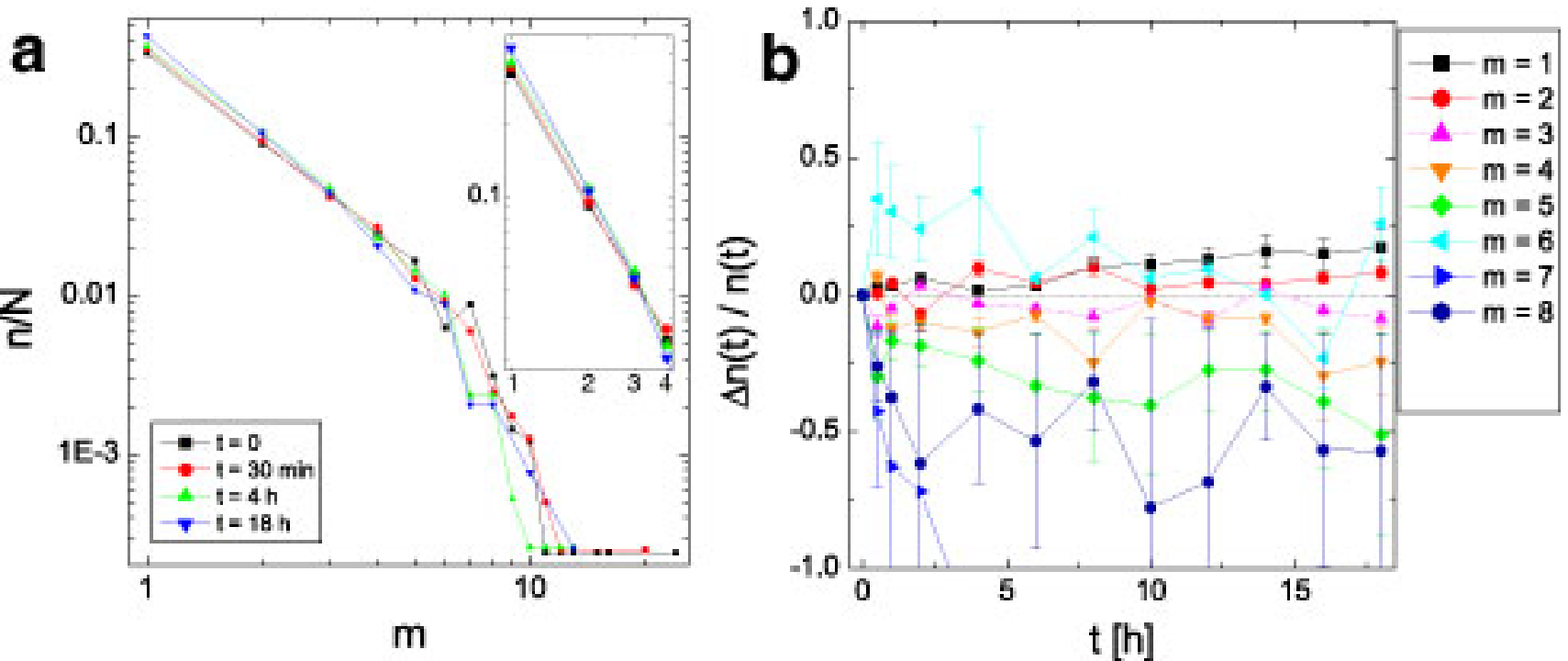}}
\noindent \textbf{SFig. 3: Cluster distribution and aging mechanism.} 
\textbf{a,} shows the variation in cluster
size distribution over time. The inset shows the distribution 
for clusters of size
$m<5$. $n$ denotes the number of clusters, $N$ the number of coordinates
(particles) found. $m$ gives the size of clusters. \textbf{b,} shows
the evolution of different cluster sizes, where $\Delta n(t)=n(t)-n(0)$.
Note the increase in the population of $m=6$ at early times.

\vspace{0.5cm}
\noindent \textbf{On the absence of mesoscopic ordered structures.}
Here we note that a considerable literature exists, predicting
pattern-formation and lamellae-like phases in systems with competing
interactions \cite{tarzia2006,groenewold2001}. 
We see little periodicity in the
system studied here. In the case of the cluster glass, this is likely
due to extremely slow equilibration to a monodisperse cluster phase
which could then potentially form a cluster crystal \cite{mladek2006}.
Our experiments run for up to two weeks, and there was 
no indication of crystalline ordering, moreover the cluster 
distribution was still very polydisperse on this timescale. 
The behavior up to 18 hours is shown in SFig. 3.

In the gel, on the other hand, we believe that
the bicontinuous network is slow to break and re-arrange due to the
connectivity \cite{tanaka2007}, as would be required to form a lamellar
type structure. Another important difference in comparing the gel
state with for example, microemulsions, lies in the arrested nature
of the `colloid'-rich phase. Microemulsions are thus able to reorganize
locally in a way that a system with short-ranged attractions, is not.

\vspace{0.5cm}
\noindent \textbf{Dependence of the state diagram 
of multiple dynamic arrested states upon the strength of electrostatic interactions.} 
The state diagram of multiple dynamic arrested states for varying strengths
of the electrostatic interactions is shown in SFig. 4. 
Upon reducing the strength of the electrostatic interaction, 
one expects to recover the behavior of hard spheres with 
a short-ranged attractive interaction
\cite{pham2002}. That is to say, fluid, gel, attractive and hard 
sphere glasses.
Here we discuss the differences this phase diagram relevant 
to neutral colloids to the strongly charged case presented here.
Note that we consider the case of no added salt such that the 
electrostatics are unscreened.

The effect of increasing electrostatic repulsions is as follows. 
When the colloid
charge $Z$ is small, a fluid (blue), gel, attractive glass (yellow)
and repulsive glass (red) states are found \cite{pham2002}. Our study
suggests that increasing the electrostatic repulsions leads to the
system studied here, with a significant extension of the repulsive
glass to low densities, such that we find a direct repulsive glass-gel
transition. This is quite different from the fluid-gel transition
in the case of small $Z$. Moreover the enhancement of repulsions 
due to the electrostatics leads to an increase in the amount of 
polymer required for gelation. 

We furthermore see the emergence of a cluster
glassy state (green), which is not found for neutral systems. 
Nonergodicity stems from connectivity in the case of
the gel and caging for single particles and clusters, respectively,
for the repulsive and cluster glasses. For a high colloid volume fraction,
there may also be an attractive glass state, although such a high
volume fraction region was not investigated in this study. We emphasize
that for large $Z$ an ergodic state exists only for a very low colloid
volume fraction and most state points form non-ergodic disordered
states. It is an intriguing, fundamental question how the state diagram
evolves between these two extremes.

\vspace{0.5cm}
\centerline{\includegraphics[width=0.5\textwidth]{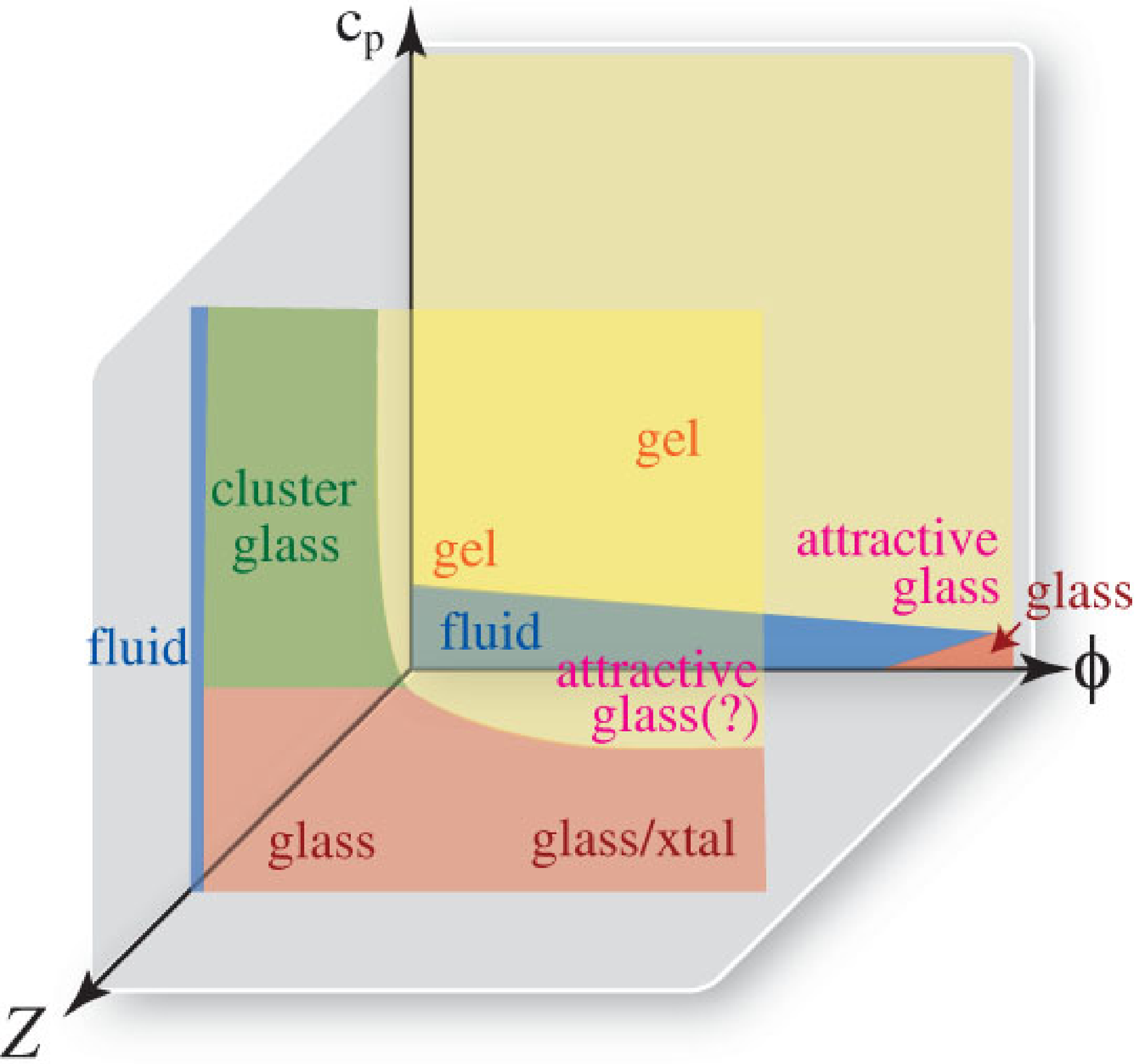}}
\noindent \textbf{SFig. 4: 3D state diagram}. The state
diagrams for two extremes, negligible and strong electrostatic interactions
($Z=0$ and large $Z$) are indicated. The former has been established
in \cite{pham2002}, while the latter is a schematic of Fig. 1. See
the text for details.


\end{document}